\documentclass[conference]{IEEEtran}
\IEEEoverridecommandlockouts
\usepackage{cite}
\usepackage{amsmath,amssymb,amsfonts}
\usepackage{algorithmic}
\usepackage{graphicx}
\usepackage{textcomp}
\usepackage{xcolor}
\def\BibTeX{{\rm B\kern-.05em{\sc i\kern-.025em b}\kern-.08em
    T\kern-.1667em\lower.7ex\hbox{E}\kern-.125emX}}
\begin{document}

\title{Detecting Sybil Addresses in Blockchain Airdrops: A Subgraph-based Feature Propagation and Fusion Approach\\
}
\author{\IEEEauthorblockN{1\textsuperscript{st} Qiangqiang Liu}
\IEEEauthorblockA{\textit{Risk Department} \\
\textit{Binance}\\
Dubai, United Arab Emirates \\
codi.l@binance.com}
\and
\IEEEauthorblockN{2\textsuperscript{st} Qian Huang}
\IEEEauthorblockA{\textit{Risk Department} \\
\textit{Binance}\\
Hong Kong, China \\
qian.huang@binance.com}
\and
\IEEEauthorblockN{3\textsuperscript{st} Frank Fan}
\IEEEauthorblockA{\textit{Risk Department} \\
\textit{Binance}\\
Hong Kong, China \\
frank.f@binance.com}
\and
\IEEEauthorblockN{4\textsuperscript{st} Haishan Wu}
\IEEEauthorblockA{\textit{AI Department} \\
\textit{Zand}\\
Dubai, United Arab Emirates \\
haishan.wu@zand.ae}
\and
\IEEEauthorblockN{5\textsuperscript{st} Xueyan Tang}
\IEEEauthorblockA{\textit{Suzhou Artificial Intelligence Research Institute} \\
\textit{Shanghai Jiao Tong University}\\
Suzhou, Jiangsu, China \\
mirror.tang@alumni.stanford.edu}
}

\maketitle

\begin{abstract}
Sybil attacks pose a significant security threat to blockchain ecosystems, particularly in token airdrop events. This paper proposes a novel sybil address identification method based on subgraph feature extraction lightGBM. The method first constructs a two-layer deep transaction subgraph for each address, then extracts key event operation features according to the lifecycle of sybil addresses, including the time of first transaction, first gas acquisition, participation in airdrop activities, and last transaction. These temporal features effectively capture the consistency of sybil address behavior operations. Additionally, the method extracts amount and network structure features, comprehensively describing address behavior patterns and network topology through feature propagation and fusion. Experiments conducted on a dataset containing 193,701 addresses (including 23,240 sybil addresses) show that this method outperforms existing approaches in terms of precision, recall, F1 score, and AUC, with all metrics exceeding 0.9. The methods and results of this study can be further applied to broader blockchain security areas such as transaction manipulation identification and token liquidity risk assessment, contributing to the construction of a more secure and fair blockchain ecosystem.
\end{abstract}


\section{Introduction}
Sybil attacks are a security threat in decentralized networks where attackers create multiple identities or nodes to influence system decision-making processes or consensus mechanisms. In computer networks and blockchain technology, the purpose of sybil attacks is typically to gain disproportionate influence or control, such as in voting systems, reputation systems, or airdrop events. In the cryptocurrency domain, sybil addresses refer to multiple addresses controlled by a single entity, used to masquerade as multiple independent participants to manipulate markets, abuse airdrop events, or influence decisions in decentralized autonomous organizations (DAOs). The existence of sybil addresses undermines the principle of decentralization in blockchain technology and compromises the fairness and trustworthiness of the system. Airdrops are a common cryptocurrency marketing strategy where projects distribute tokens to a wide range of users to increase their project's visibility and user base. However, airdrop events are also susceptible to sybil attacks, where attackers create multiple addresses to collect more airdrop tokens, unfairly increasing their gains. This behavior not only infringes on the rights of other users but also potentially damages the project's reputation and market value.

Sybil addresses have garnered increasing attention in various blockchain projects (e.g., Starknet, zkSync, LayerZero), leading to the accumulation of labeled sybil address datasets. Machine learning models and graph neural networks \cite{b7, b21} have demonstrated exceptional efficacy in identifying entity addresses and fraudulent addresses in blockchain systems. However, academic literature specifically addressing sybil address identification remains relatively scarce, with only two directly relevant works employing unsupervised methods for this purpose.  Airdrop events can be categorized into short-term and long-term distributions: 

(1) Short-term airdrops: These involve one-time token distributions by project teams on a specific dataset.

(2) Long-term airdrops: These encompass prolonged token distributions on a specific dataset, exemplified by Soulbound Token (SBT) airdrops or projects conducting multiple-phase airdrops or extended marketing campaigns.
These distinct categories necessitate different analytical approaches. For short-term airdrops, unsupervised methods are more appropriate, facilitating the exploration and identification of sybil address behavioral patterns. Conversely, long-term airdrops are better suited to supervised methods, as the extended distribution period allows for the accumulation of a substantial corpus of labeled sybil addresses, enabling more precise supervised model training and subsequent identification. 

As we amass diverse sybil address datasets across various platforms, we gain insights into a wide spectrum of sybil behavioral patterns. To achieve optimal identification efficacy, supervised methodologies emerge as the most promising approach. This strategy leverages the rich, labeled data accumulated over time, allowing for more nuanced and accurate detection of sybil addresses across different contexts and blockchain ecosystems. Inspired by Chen et al.\cite{b1}, this paper proposes a supervised subgraph-based feature extraction algorithm, subgraph-based lightGBM, to identify sybil addresses by extracting time series of critical operations and network structural features. The contributions of this study are as follows:

(1) This research presents the first application of a supervised machine learning method to sybil address identification. The proposed Subgraph-based lightGBM algorithm captures structural features of airdrop address subgraphs and extracts composite temporal features based on the sybil address lifecycle (first transaction, first gas recharge, participation in airdrop events, and last transaction time).

(2) We conduct network feature analysis on the transaction subgraphs of sybil addresses and analyze the importance and contribution of these features to reveal how they capture the transaction patterns of sybil addresses.

(3) We validate the proposed model's effectiveness in describing sybil address transaction patterns using a dataset containing 193,701 addresses (including 23,240 sybil addresses). Our results demonstrate that our model outperforms the existing Trusta model.

The structure of this paper is organized as follows: Section~\ref{sec:2} presents a comprehensive literature review, encompassing relevant works on entity identification and sybil address detection in blockchain systems. Section~\ref{sec:3} introduces our proposed Subgraph-based lightGBM algorithm, detailing its methodology and implementation. Section~\ref{sec:4} discusses our experimental results, including feature analysis and feature interpretability, demonstrating the efficacy of our approach. Section~\ref{sec:5} addresses the limitations of our study, providing a critical evaluation of our methodology and results. Finally, Section~\ref{sec:6} concludes the paper with a summary of our findings and offers perspectives on future research directions in this domain.

\section{Related Works} \label{sec:2}

\subsection{Entity Identification}
Entity identification, which associates multiple account addresses belonging to the same entity, has become a crucial step in revealing the real identities behind blockchain accounts and provides an important technical basis for sybil address detection. Existing entity identification methods are mainly divided into three categories: transaction-based, behavior-based, and off-chain information-based, each with its own characteristics, aiming to parse blockchain data from different angles to reveal entity relationships behind complex transaction networks. Wu et al.\cite{b2} and Xu et al.\cite{b3} used graph neural networks and graph attention networks, respectively, to improve identification accuracy. Cross-chain analysis has also made progress, with Victor and Lüders\cite{b4} proposing a cross-chain entity association method based on address reuse patterns, and Zhang et al.\cite{b5} developing a multi-modal learning framework. Jourdan et al.\cite{b6} studied the use of machine learning techniques to track cross-chain transactions. The privacy-preserving entity identification framework based on federated learning proposed by Li et al.\cite{b7}, and the privacy-preserving anti-money laundering analysis method discussed by Biryukov and Tikhomirov\cite{b8}, both demonstrate the balance between privacy and security. The research by Tzanetakis et al.\cite{b9} showcases the potential application of entity identification in regulating illegal markets.

Overall, blockchain entity identification research has developed to a stage capable of handling complex network structures, conducting cross-chain analysis, and considering privacy protection. However, sybil address identification has its own specificity, focusing more on identifying malicious actors, requiring stricter judgment criteria and higher identification accuracy. Sybil addresses typically exhibit more consistent or coordinated behavior patterns and may appear concentrated within specific time periods. Therefore, when applied to sybil address identification, special attention needs to be paid to these characteristics and the constantly changing strategies of attackers.

\subsection{Sybil Address Detection}
In blockchain security research, while sybil attacks are a widely discussed topic, studies specifically targeting sybil address identification at the transaction level, particularly in airdrop scenarios, remain relatively scarce. Existing relevant research primarily focuses on address clustering, anomalous transaction detection, and network structure analysis. Although these methods are not directly aimed at sybil address identification, they provide potential technical foundations for addressing this issue.

The Ethereum address clustering method proposed by Victor \cite{b10}, while primarily intended to identify addresses belonging to the same entity, could potentially be applied to identify multiple sybil addresses controlled by the same attacker. Payette et al.'s\cite{b11} feature analysis of the Ethereum address space provides a basis for understanding normal and abnormal address behaviors, which may aid in distinguishing between regular users and sybil addresses. In terms of anomalous transaction detection, although the studies by Hu et al.\cite{b12} and Farrugia et al.\cite{b13} focus on Bitcoin money laundering activities and illegal Ethereum accounts respectively, their proposed methods could potentially be adapted to identify abnormal participation behaviors in airdrops. Network analysis methods, such as those in the research of Wu et al.\cite{b14} and Lin et al.\cite{b16}, while mainly concerned with fraud detection and general transaction pattern analysis, could potentially apply their network embedding and complex network analysis techniques to identify collaborating groups of sybil addresses. The entity characterization method proposed by Jourdan et al.\cite{b15}, although conducted on the Bitcoin network, could potentially be extended to identify anomalous entities in airdrops. Toyoda et al.'s\cite{b18} analysis of transaction patterns in high-yield investment programs, despite the different context, might offer insights into identifying abnormal participation patterns in airdrops. Chauhan et al.\cite{b17} discussed blockchain scalability issues which, while not directly related to sybil address identification, provide valuable reference for handling large-scale transaction data and constructing efficient identification systems. Baumann et al.\cite{b19} explored the Bitcoin network relatively early, laying the groundwork for subsequent network analysis and anomaly detection research. Although these studies do not directly target sybil address identification in airdrop scenarios, they provide methods and techniques that could potentially be applied or improved to address this issue.

\begin{figure}[htbp]
    \centering
    \includegraphics[width=\columnwidth]{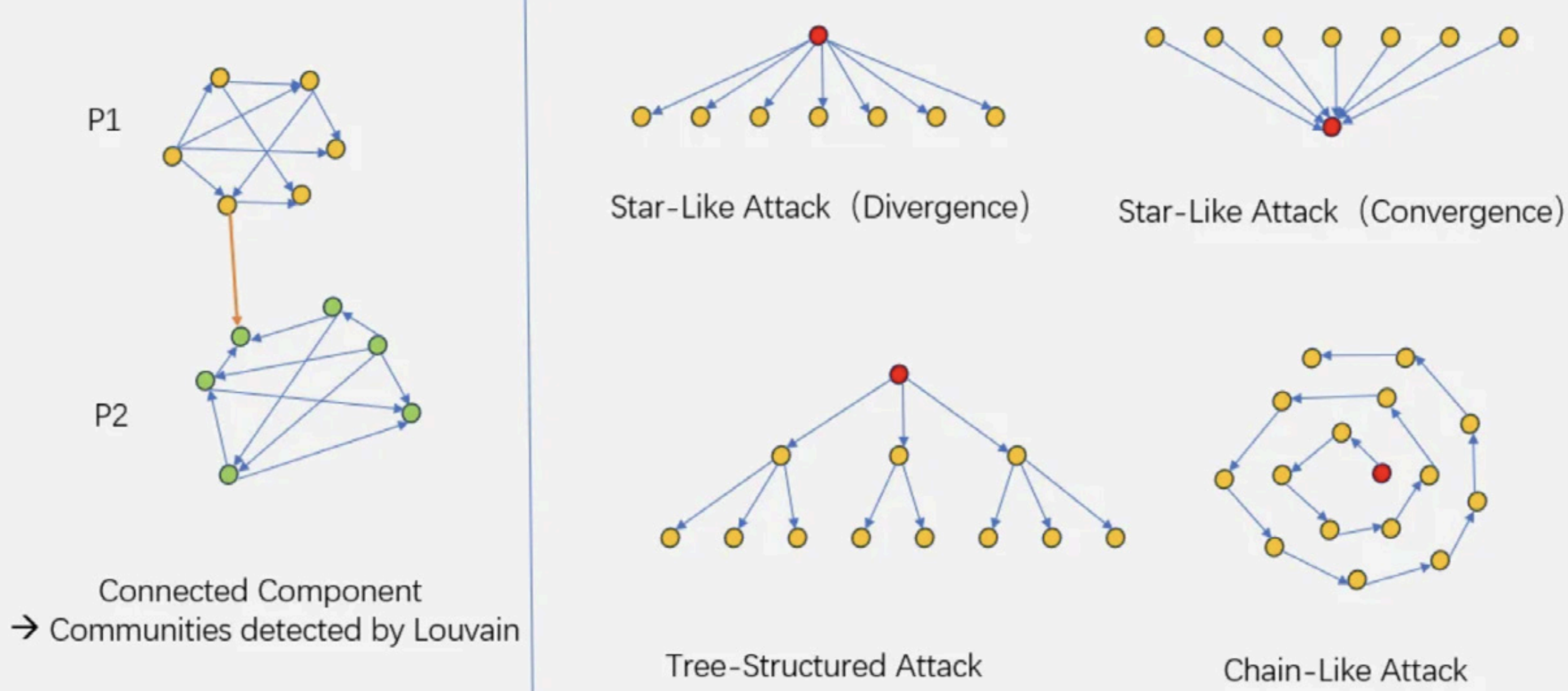}
    \caption{Community detection on asset transfer graphs (ATGs), via GitHub (https://github.com/TrustaLabs/Airdrop-Sybil-Identification).}
    \label{fig:trusta}
\end{figure}

As shown in Fig.~\ref{fig:trusta}, Trusta and Liu et al.\cite{b20} first use graph mining algorithms to identify sybil addresses, then employ user analysis to filter outliers and improve precision. This approach requires case analysis for each airdrop event, while the method has broad applicability, it demands substantial human effort to identify transaction patterns and adjust clustering parameters.



\section{Methodology} \label{sec:3}
Current sybil address identification methods primarily focus on unsupervised approaches, extracting sybil addresses by analyzing transaction graphs constructed from address fund sources, destinations, transaction networks, and frequencies. The advantage of this approach is that it doesn't require labeled sybil datasets and offers a universal methodology applicable to almost all scenarios. However, it demands significant human effort. The challenges currently faced in sybil address identification are as follows: 

(1) Sybil addresses typically exhibit similar behavioral patterns and may appear concentrated within specific time periods. Moreover, attackers' strategies are constantly evolving. 

(2) Constructing transaction graphs requires tracing multiple layers of transaction relationships. During this process, some addresses like hot wallets or contracts generate many transactions, resulting in enormous transaction graphs that require substantial computational resources. Additionally, some entity addresses may interfere with the results. 

(3) Current mainstream unsupervised sybil address identification methods, after identifying sybil transaction pattern graphs, still need to use clustering to exclude addresses with significantly different behavioral patterns. The clustering process involves selecting clustering thresholds. Threshold confirmation is challenging, often based on vague empirical choices, and the results cannot be verified.

To address challenge 1, in Section~\ref{sec:3.1}, we extract key event time-series features according to the sybil address lifecycle (first transaction, first gas recharge, participation in airdrop events, last transaction time). These features can describe the consistency of sybil address behavioral operations. To address challenges 2 and 3, in Section~\ref{sec:3.2}, we propose a supervised Subgraph-based lightGBM algorithm. This method constructs transaction subgraphs of sybil addresses, reducing graph scale and computational load by avoiding full graph data expansion. It also extracts structural features of sybil address transaction subgraphs to describe their subgraph morphology.

\subsection{Features Extraction} \label{sec:3.1}
The only data available for our analysis is the transaction data of addresses, which includes transaction\_hash, input\_address (sender address), output\_address (recipient address), network, coin (transaction currency), amount (native token amount), amount\_usdt (USDT value of the transacted token), gas\_fee, and transaction\_time. Sybil addresses tend to exhibit temporal clustering, for instance, one address distributing similar amounts of tokens to many of its other addresses to participate in certain activities. This observation underscores the importance of time and amount features. We have extracted three categories of features: time features, amount features, and transaction network structure features. These features are described in detail as follows:

\begin{itemize}
\item Time Features
    \begin{itemize}
    \item Time of first gas fee receipt: This is the timestamp of the first transaction where the address received native tokens as a recipient (native tokens refer to, for example, ETH on the Ethereum network). Every time an address initiates a transaction (as a sender), it needs to pay a gas fee.
    \item Time of first transaction: The timestamp of the address's first transaction. This may coincide with the time of first gas fee receipt.
    \item Time of first activity participation: To receive token airdrops, addresses need to participate in certain interactive activities set by the project.
    \end{itemize}
\item Amount Features: Amount features include the address balance and transaction amounts. Transaction amounts can be further categorized into amounts sent and amounts received by the address.
\item Transaction Network Structure Features
    \begin{itemize}
    \item Address degree: in-degree and out-degree.
    \item Neighbor information: number of inflow/outflow addresses in the first/second layer of transactions originating from the address.
    \item Network/coin information: number of coin types and networks (blockchain networks) the address has transacted on.
    \end{itemize}
\end{itemize}

These features are specifically designed to capture the characteristic behaviors of sybil addresses. Time features reveal suspicious patterns where sybil addresses often display abnormally tight temporal clustering—they are typically created shortly before airdrops, with minimal intervals between receiving gas, conducting first transactions, and participating in qualifying activities. This "just-in-time" pattern differs significantly from legitimate users who create addresses for long-term use, resulting in natural, irregular intervals between these events.

Transaction network structure features are particularly effective at identifying star topology patterns, which strongly correlate with sybil activity in blockchains. In a star topology, a central controlling address (hub) distributes assets to multiple controlled addresses (spokes), manifesting as abnormally high out-degree for the hub and limited transaction history beyond hub interactions for the spokes. This topology represents a cost-efficient strategy for attackers to maintain multiple addresses while minimizing transaction fees, and our extraction methodology effectively captures these coordinated operations.

These features can be combined in various ways. For time features, we can extract: the interval between the first transaction time and the first gas fee receipt time, the interval between the first activity participation time and the first transaction time, the interval between the first activity participation time and the first gas fee receipt time, and the address's active time (interval between the last and first transaction times). For amount features, we can extract the minimum, maximum, average, median, and variance of the amounts.

\subsection{Subgraph-based Feature Extraction, Fusion, and Transfer Method} \label{sec:3.2}

\begin{figure*}[htbp]
    \centering
    \includegraphics[width=0.8\textwidth]{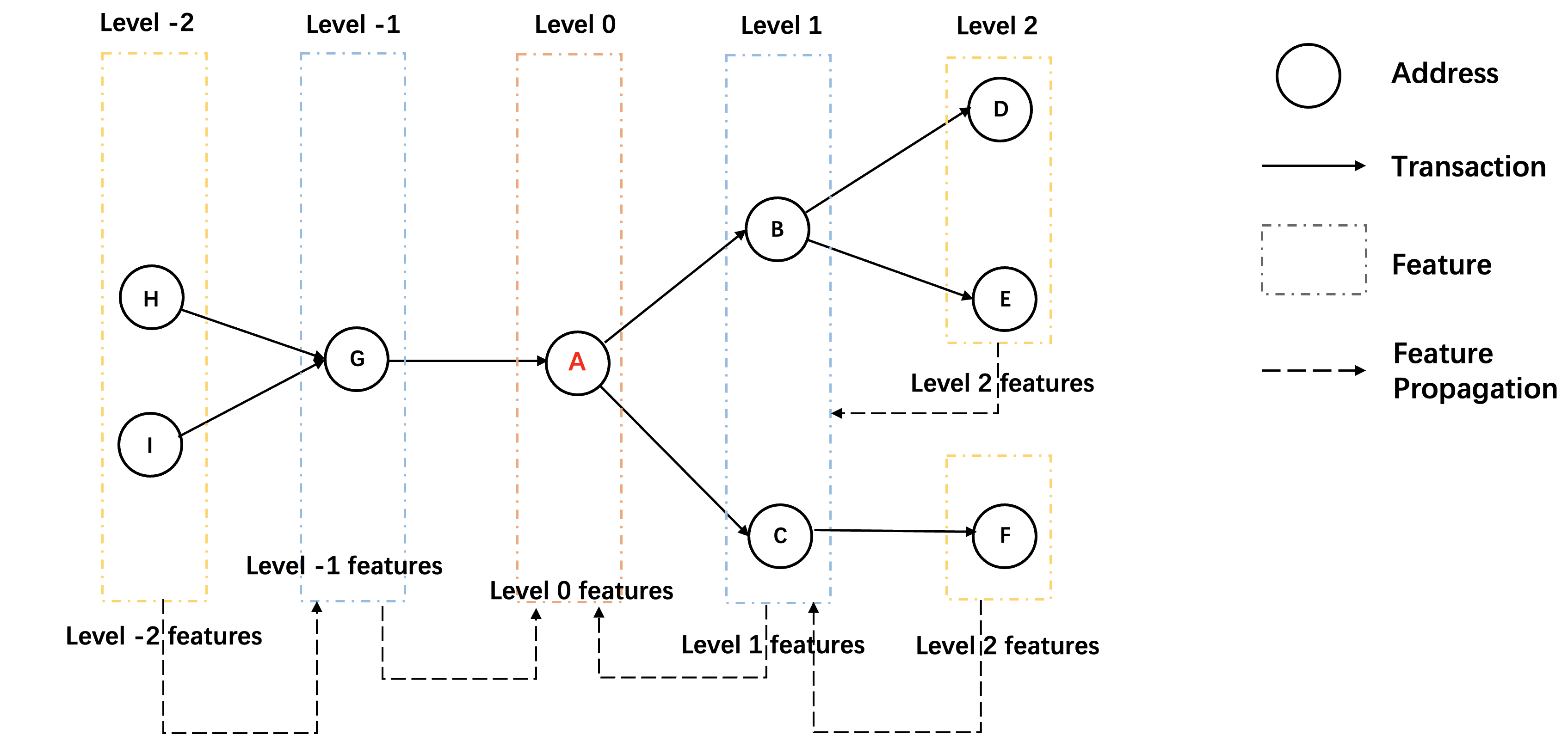}
    \caption{Example of 2 levels features fusion.}
    \label{fig:features}
\end{figure*}

Graph-based features have been proven effective by Ramalingam et al.\cite{b22}. As shown in Fig.~\ref{fig:trusta}, transaction subgraphs of sybil addresses often exhibit star-shaped, chain-shaped, and tree-shaped patterns, highlighting the importance of network features in transaction graphs. Inspired by Chen et al.\cite{b22}, we use cascading features based on transaction graphs to extract structural features. Let $TG = (V, E)$ be a transaction graph where $V$ represents addresses and $E = \{(v_i, v_j)|v_i, v_j \in V\}$ represents transactions. For a target address $A \in V$ at level 0, we define:
$\mathcal{F}_n(A) = \text{ set of n-order friends of } A$. The cascade feature extraction process consists of three steps:

(1) Layer Extraction: As shown in Fig.~\ref{fig:features}, for a sample address A, we extract two layers of transactions above and below it through transaction relationships. For example, address A at Level 0 transfers to addresses B and C at Level 1, while address G at Level -1 transfers to address A.

   $\mathcal{L}(A) = \mathcal{F}_{-2}(A) \cup \mathcal{F}_{-1}(A) \cup \{A\} \cup \mathcal{F}_1(A) \cup \mathcal{F}_2(A)$

 (2) Feature Computation: We extract time features, amount features, and transaction network structure features for addresses at each level, as described in Section~\ref{sec:3.1}. For A's inflow layers (Level -2 and -1), we extract inflow features for each address. For A's outflow layers (Level 1 and 2), we extract outflow features for each address. For each level $l$: 
 
   $\mathcal{T}_l = \{t_1, t_2, ..., t_7\}$ (time features)
   
   $\mathcal{A}_l = \{a_1, a_2, ..., a_{60}\}$ (amount features)
   
   $\mathcal{N}_l = \{n_1, n_2, ..., n_8\}$ (network features)

(3) Feature Propagation and Fusion: We propagate and fuse features from each layer towards address A at Level 0. For example, features from addresses D and E at Level 2 are propagated to address B at Level 1, then features from B, D, and E are propagated to A. Using the amount feature as an example, we merge the outflow transaction amount arrays for addresses A, B, D, and E, then calculate statistics like minimum, maximum, average, and variance. Different types of features undergo different operations; for instance, degree features are summed. Time features are calculated directly from A's transactions without propagation. 

    For amount features: 
    
   $\Phi_A(\{a_i^j\}_{j=1}^k) = \{min, max, avg, var\}(\{a_i^j\}_{j=1}^k)$
   
   For degree features: 
   
   $\Phi_D(\{d_i^j\}_{j=1}^k) = \sum_{j=1}^k d_i^j$

The final feature set:

$\mathcal{F}(A) = \mathcal{T}_0 \cup \Phi(\bigcup_{l=-2}^2 \mathcal{A}_l) \cup \Phi(\bigcup_{l=-2}^2 \mathcal{N}_l)$

In total, we extract 75 features: $7 \text{ (time features)} + 2 \times 6 \times 5 \text{ (amount features)} + 8 \text{ (network features)}$. In summary, for each sample address A, we first extract its transaction subgraph encompassing two layers above and below, then obtain features for address A through feature propagation and fusion. These features of address A represent the characteristics of A's transaction subgraph and can be considered as the subgraph's feature representation.

\section{Experiments} \label{sec:4}
\subsection{Data Collection and Preparation}

The labeled sybil address data was collected through a rigorous data quality validation process from the BAB (Binance Account Bound) airdrop event. BAB is Binance's first Soul Bound Token launched in 2022, designed as a non-transferable token on BNB Smart Chain to verify users who completed Binance's KYC process, linking on-chain activities with real identities. We initially identified suspected sybil addresses through comprehensive manual analysis and clustering methods. After identifying these suspicious addresses, we reclaimed rewards from them and established a manual appeal review process. Only addresses that did not appeal or had their appeals rejected after thorough manual review were classified as confirmed sybil addresses.

For data preparation, we performed meticulous cleaning on the transaction records. We queried address labels for all involved addresses using Arkham Intelligence's API interface. We selected Arkham Intelligence (https://intel.arkm.com/) for our research because it is a leading blockchain analytics platform offering comprehensive address labeling and transaction tracking services across multiple blockchains, with extensive capabilities for identifying institutional wallets, smart contracts, and related address clusters. The cleaning process involved: 1) Excluding institutional addresses, including hot wallet addresses and contract addresses, as these are unlikely to be sybil addresses. 2) Retaining transactions distributed by contracts, as this represents a common sybil address pattern where funds are sent to multiple addresses in a single transaction through contract distribution protocols. 3) Filtering out addresses with lifecycles exceeding one year (accounting for 2.6\% of the dataset). Since creating addresses on-chain has minimal cost, sybil attackers typically exhibit high address abandonment rates to evade detection.


Unlike Chen et al.\cite{b1}, we did not filter based on amount or transaction frequency. This decision was made considering that addresses with small amounts might be used for gas distribution\cite{b23}, addresses with high transaction frequencies are more likely to be sybil addresses, and those with low transaction frequencies might serve as connecting addresses between sybil groups. After this filtering process, our dataset comprises 193,701 addresses (including 23,240 sybil addresses). The data spans from January 2023 to May 2024. By extracting two layers of transaction data above and below these addresses, we encompassed a total of 58,397,048 transactions.

\begin{table}[htbp]
\caption{The performance comparison}
\label{tab:comparison}
\begin{center}
\begin{tabular}{|c|c|c|c|c|}
\hline
\textbf{Method} & \textbf{Precision} & \textbf{Recall} & \textbf{F1} & \textbf{AUC} \\
\hline
SVM & 0.6203 & 0.0329 & 0.0624 & 0.5599 \\
\hline
DT & 0.7289 & 0.7572 & 0.7428 & 0.7373 \\
\hline
lightGBM & 0.7821 & 0.7150 & 0.7470 & 0.8484 \\
\hline
Clustering-based Trusta & 0.7962 & 0.8159 & 0.8059 & 0.8642 \\
\hline
\textbf{Subgraph-based lightGBM} & \textbf{0.9428} & \textbf{0.9182} & \textbf{0.9303} & \textbf{0.9806} \\
\hline
\end{tabular}
\end{center}
\end{table}

\subsection{Base Model}

To demonstrate the effectiveness of our proposed Subgraph-based lightGBM model, we compared it with classical machine learning models: lightGBM, Support Vector Machine (SVM), and Decision Tree (DT). We also compared it with the Clustering-based sybil address identification model from Trusta, a leading company in this field. LightGBM is particularly important for our approach as it efficiently handles complex feature interactions and captures non-linear relationships between on-chain activities - crucial for detecting subtle sybil behavior patterns. Its histogram-based approach and Gradient-based One-Side Sampling (GOSS) allow us to effectively identify and utilize the most informative features while minimizing computational overhead, making it ideal for processing large-scale blockchain transaction data.

Brief descriptions of these models are as follows: 1) Support Vector Machine (SVM): A supervised learning algorithm for classification and regression, which finds the hyperplane maximizing the margin between classes. 2) Decision Tree (DT): A simple, interpretable supervised learning algorithm that models decisions as a tree structure. 3) LightGBM: A gradient boosting framework designed for speed and efficiency, using a histogram-based algorithm to reduce data points. 4) Clustering-based Trusta: A two-stage AI and machine learning framework for sybil address identification. It first uses community detection algorithms to analyze asset transfer graphs (ATG) and identify suspicious densely connected groups. Then, it refines these groups using K-means clustering based on user profiles and activities. For SVM, DT, and LightGBM, we used first-order features mentioned by Farrugia et al.\cite{b24} for model training. For Clustering-based Trusta, time features were prioritized during clustering.

\subsection{Experiment Result and Analysis}

To validate the effectiveness of our proposed model, we compared SVM, DT, lightGBM, Clustering-based Trusta, and our Subgraph-based lightGBM model. As shown in Table~\ref{tab:comparison}, our model demonstrates superior performance across all evaluation metrics.
The Precision of traditional models is too low for practical requirements, leading to numerous False Positives that can damage project reputation. In terms of Recall, SVM performs poorly at 0.0329, while other models show moderate values between 0.7150 and 0.8159. Our Subgraph-based lightGBM achieves a recall of 0.9182, identifying nearly all sybil addresses.
The F1 score confirms our model's balanced performance with 0.9303, compared to SVM's low 0.0624 and traditional models' moderate scores between 0.7428 and 0.8059. For AUC, our model achieves 0.9806, significantly outperforming others, with Clustering-based Trusta next at 0.8642.
Notably, our proposed model achieves scores exceeding 0.9 across all metrics, fully meeting real-world application requirements where both precision and comprehensive detection are critical.

Our comparative analysis of transaction patterns in Fig.~\ref{fig:trusta} demonstrates our model's enhanced efficacy in sybil address detection. For 'Star' shaped patterns, where Trusta classifies addresses as High, Medium, or Low Risk, our model identifies problematic instances with probabilities of 99\%, 95\%, and 65\% respectively. In 'Chain' shaped patterns, classified by Trusta as High or Medium Risk, our model detects issues with 100\% and 95\% probability respectively. Similarly, for 'Tree' shaped patterns, also categorized as High or Medium Risk by Trusta, our model identifies problematic addresses with 97\% and 95\% probability respectively. The high concordance with Trusta's high-risk classifications, coupled with our model's ability to refine medium and low-risk categorizations, demonstrates both improved precision and recall in sybil address detection.

\begin{figure}[htbp]
    \centering
    \includegraphics[width=\columnwidth]{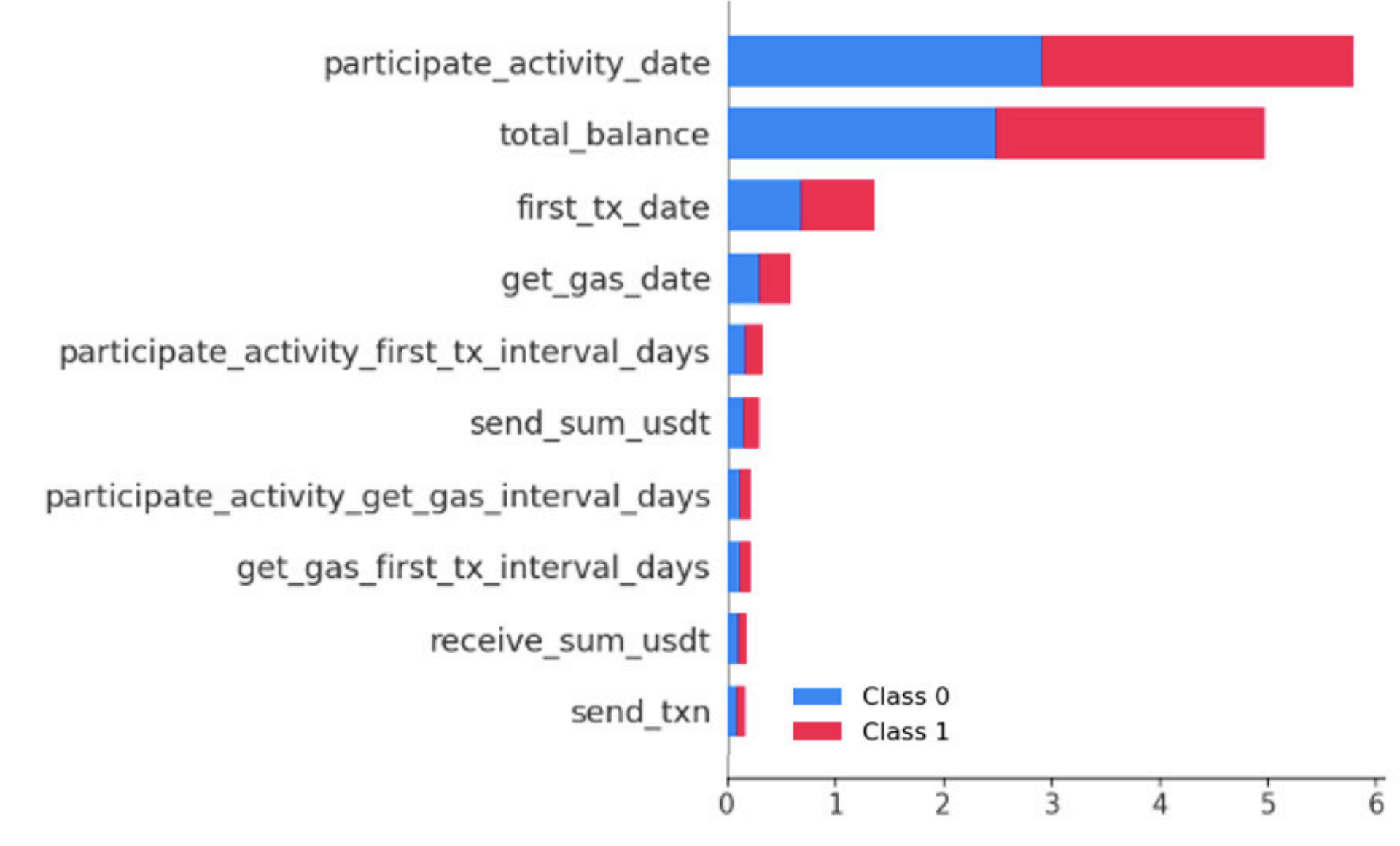}
    \caption{The top 10 important features.}
    \label{fig:features_importance}
\end{figure}

Fig.~\ref{fig:features_importance} illustrates the top 10 most important features extracted using our subgraph-based method. Key findings from our analysis include:

\begin{itemize}
\item participate\_activity\_date/first\_tx\_date/get\_gas\_date: These three features represent the time of first participation in an airdrop activity, first transaction, and first gas fund receipt, respectively. For sybil addresses, there is a notable temporal clustering in their operational behaviors. This demonstrates that our extracted key operational time features can effectively describe the behavioral consistency of sybil addresses, addressing challenge 1 from Section~\ref{sec:3}.
\item total\_balance/send\_sum\_usdt: Typically, sybil addresses maintain a balance slightly above the minimum requirement for participating in airdrop activities. This strategy aims to minimize capital costs, allowing the creation of more sybil addresses. After receiving the airdrop, funds are usually transferred out and the address is abandoned.
\end{itemize}

\section{Limitation and Discussion} \label{sec:5}
Despite the promising results achieved by the Subgraph-based lightGBM model in sybil address identification, there remains room for improvement: 1) Model Generalizability: The current model relies on features specific to certain datasets, which limits its direct applicability to new datasets. However, the critical event sequence features and structural features we extract possess inherent universality and generalization capacity. Unsupervised methods can be employed for preliminary labeling of new datasets before applying the supervised model. Our proposed methodology is particularly suited for the long-term airdrop events discussed in the introduction. 2) Model Applicability: While the model demonstrates robust performance on account-based blockchains and can be relatively easily extended to other EVM-compatible chains, it is not applicable to UTXO-based blockchains. This constraint limits the model's widespread application. However, it's worth noting that due to high gas fees, airdrop events are generally not conducted on UTXO-model blockchains, mitigating this limitation in practice.

\section{Conclusion and Future Work} \label{sec:6}

In cryptocurrency airdrops, sybil addresses pose a significant challenge, with some notable events containing over 30\% sybil addresses. These addresses misuse project resources and prevent fair token distribution to genuine participants. Therefore, identifying sybil addresses is crucial for maintaining fairness in decentralized systems.

This paper introduces a Subgraph-based lightGBM algorithm that captures structural features of airdrop address subgraphs and extracts key event sequence features based on the sybil address lifecycle. Our approach focuses on universal behavioral patterns rather than project-specific features, making it adaptable to various airdrop scenarios. We validated our method using the BAB (Binance Account Bound) airdrop dataset with a comprehensive data quality validation process.

While we compared our method with basic models and the existing Trusta sybil address identification model, we acknowledge the need for broader comparisons in future work, including advanced graph-based models like GCN, GAT, and TGNN. Our subgraph feature extraction approach provides better interpretability and lower computational overhead compared to traditional methods.

Future work will extend this research to more networks and airdrop events, establishing a comprehensive sybil address database with an iterative improvement cycle for model updates. Sybil address identification is foundational for various on-chain research areas beyond airdrop distribution, such as identifying transaction manipulation and token liquidity risks.



\end{document}